%% file: rs-arxiv.tex
\documentclass[conference]{IEEEtran}
\usepackage[T1]{fontenc}
\usepackage{cite}
\usepackage{enumitem}
\usepackage{amsmath}
\usepackage{graphicx}
\usepackage{listings}
\usepackage{url}
\usepackage{hyperref}
\usepackage[nameinlink]{cleveref}
\usepackage{algorithm,algorithmicx}
\usepackage[noend]{algpseudocode}
\usepackage[font=footnotesize,labelsep=period,skip=3pt]{caption}
\usepackage[font=footnotesize]{subcaption}
\usepackage{xcolor}
\usepackage{tikz}
\usetikzlibrary{arrows.meta,positioning,shapes.geometric,calc}

\renewcommand{\theparagraph}{%
  \ifnum\value{subsubsection}=0\thesubsection(\alph{paragraph})%
  \else\thesubsubsection\alph{paragraph}\fi}

\makeatletter
\def\fps@figure{tbp}
\makeatother

\hypersetup{
  pdftitle={HEL: A Heterogeneous Extensible Layer for GPU-Resident Runtime Orchestration},
  pdfauthor={Agnieszka M. \L{}upi\'nska, Serban D. Porumbescu, John D. Owens}
}

\begin{document}

\title{HEL: A Heterogeneous Extensible Layer for\\GPU-Resident Runtime Orchestration}

\author{%
  \IEEEauthorblockN{Agnieszka M. \L{}upi\'nska, Serban D. Porumbescu, and John D. Owens}
  \IEEEauthorblockA{University of California, Davis\\Davis, California, USA}
}

\maketitle

\begin{abstract}
\input{rs/abstract}
\end{abstract}

\input{rs/intro}
\input{rs/background}
\input{rs/design}

\input{rs/programming_model}
\input{rs/implementation}

\input{rs/tests}

\input{rs/case_study}
\input{rs/conclusions}
\input{rs/acknowledgement}

\bibliographystyle{IEEEtran}
\bibliography{rs}

\end{document}

%% file: rs/abstract.tex
We present HEL, a modular GPU-resident runtime whose stateful services persist across application kernel launches.
Application programmers call these services from device code.
Service developers express coroutine-style computations as dependency-linked tasks managed by HEL.
HEL supports synchronous and asynchronous calls, runtime dependencies, batching, and automatic retries. 
Application kernels and service tasks execute concurrently on spatially partitioned GPU resources. 
Producer-Consumer and Host-RPC microbenchmarks show gains in underutilized and latency-hiding regimes, while a PFPL-based Checkpoint Compression case study achieves up to $1.19\times$ application-path speedup.

%% file: rs/intro.tex
\section{Introduction}%
\label{sec:intro}

Modern GPU workloads---such as scientific simulations, data-science pipelines, and ML inference---increasingly execute multi-stage pipelines with data-dependent coordination rather than isolated kernels.
Although GPUs support dynamic work generation, device-initiated I/O, and asynchronous execution, stage coordination remains largely CPU-driven via host loops that launch kernels, synchronize results, and schedule subsequent work.
Host-driven coordination introduces launch and synchronization overhead and limits latency hiding for fine-grained stages.
As GPU parallelism grows, sustaining utilization in irregular pipelines requires reusable orchestration across phases, not just faster kernels.

Embedding orchestration within application kernels couples runtime logic to application execution, complicating its reuse across workloads and persistence across kernel launches.
Although persistent kernels~\cite{persistentThreads} keep orchestration on the GPU, combining application and runtime logic forces application kernels into a persistent-worker loop and tightly couples both layers.
The challenge is providing persistent, reusable GPU runtime functionality without imposing a persistent-kernel structure on applications.

Our central question is twofold: \emph{can GPU runtime functionality be provided as a reusable, device-invoked service, and when do persistence and concurrent execution justify its resource and coordination costs?}
For HEL-managed work, the \emph{compute plane} comprises user-defined task computations, while the \emph{control plane} schedules tasks according to their dependencies and manages their runtime state.
HEL (Heterogeneous Extensible Layer) provides this orchestration through a persistent GPU-resident runtime whose workers also execute the submitted task computations.
Application kernels and HEL execute concurrently on spatially partitioned GPU resources.
HEL workers, queues, and state persist across application kernel launches.

This design cleanly separates \emph{using} a GPU service from \emph{implementing} one.
Application kernels invoke services asynchronously, continue independent computation, and use task handles (Section~\ref{sec:programming-model}) to express dependencies or wait for results when needed.
Service developers implement reusable services using C++ task lambdas executed at thread, warp, or block granularity.
These tasks can spawn further work and express dependencies, while HEL provides persistent workers and shared control-plane machinery.
For example, an application kernel can request compression of a completed buffer and continue independent computation.
The service developer supplies the compression task.
HEL schedules and executes it, tracks its completion, and makes dependent tasks runnable.
As a stepping stone toward GPU-resident OS services, HEL could potentially support capabilities separable from kernel computation, reusable or stateful across launches, and latency-tolerant or irregular---from dynamic memory allocation with garbage collection to GPU-initiated inter-node communication for asynchronous halo exchange.

We implement a CUDA prototype of HEL and evaluate it on a single GPU across Producer-Consumer and Host-RPC microbenchmarks and a Checkpoint Compression case study.
We vary orchestration pressure and compute intensity to characterize control-heavy, compute-heavy, and latency-hiding regimes and quantify the tradeoff between coordination cost and computation. This paper makes the following contributions:
\begin{itemize}[leftmargin=*, nosep, topsep=2pt]
    \item \textbf{HEL Runtime Architecture}: A system that shifts the GPU control plane to a persistent, GPU-resident service capable of coordinating application tasks across kernel boundaries.
    \item \textbf{Task-based Orchestration Interface}: An asynchronous device-side service-call interface with explicit dependency tracking, allowing service computations to be decomposed into runtime-managed tasks.
    \item \textbf{Multi-Granularity Task Dispatch}: A runtime implementation supporting thread-, warp-, and block-granularity tasks to match dispatch granularity to the workload.
    \item \textbf{Persistent GPU Service Model}: Long-lived runtime services executing concurrently with kernels via spatial partitioning to overlap coordination and compute.
    \item \textbf{System Characterization}: A CUDA evaluation across control-heavy and latency-hiding regimes, plus a Checkpoint Compression case study, quantifying tradeoffs between coordination overhead and compute intensity.
\end{itemize}

%% file: rs/background.tex
\section{Background and Prior Work}%
\label{sec:background}%
\label{sec:prior-work}

\paragraph{HEL is fundamentally different than prior GPU ``runtime'' work}
Because ``runtime'' is used broadly, we emphasize HEL's core distinction: whereas prior GPU runtimes are primarily host-side (CPU) workload managers, HEL runs completely on dedicated GPU resources as a persistent system service---effectively a GPU microkernel---achieving low-latency autonomy unattainable from the host.
Unlike CPU-orchestrated runtimes (StarPU~\cite{Augonnet:2011:SAU}, Legion~\cite{Bauer:2012:LEL}, PaRSEC~\cite{Bosilca:2013:PEH}), HEL keeps its entire control plane on the GPU, avoiding PCIe and CPU scheduling overheads.
Whereas OpenMP~\cite{OpenMP:2024:OAP} and IRIS~\cite{Kim:2021:IAP} use a host-managed ``push'' queue, HEL implements a ``resident-pull'' model that autonomously schedules tasks onto spatially partitioned SMs.
CUDA Graphs also support device-side launches and conditional execution~\cite[Sec.~6.2.8.7]{NVIDIA:2025:CUDA}; HEL instead exposes a resident task service whose queues and state span application launches.
Finally, unlike kernel-internal persistent schedulers such as Atos~\cite{Chen:2022:AAT} and GTaP~\cite{Maeda:2026:GAG} that share one kernel's context temporally, HEL runs as a utility service on spatially partitioned GPU resources for one or more applications.

\paragraph{Coroutines, Task Programming}
Coroutines suspend and resume computation while preserving execution state~\cite{moura2009revisiting}, decomposing work into a task dependency graph scheduled by precedence rules.
CUDA C++ does not support coroutines in device code~\cite{NVIDIA:2025:CUDA}.
While Zheng et al.~\cite{Zheng:2024:GCF} introduce kernel-level GPU coroutines and schedulers for irregular rendering tasks, their approach lacks a persistent service runtime for overlapped orchestration.
HEL instead enables coroutine-style decomposition without compiler support by exposing an explicit task-and-dependency interface whose state and task graph are managed on the GPU by a persistent service.

\paragraph{GPU Spatial Partitioning}
Unlike temporal multitasking, which time-slices the entire GPU across processes or kernels, spatial partitioning~\cite{spatialMultitasking} executes concurrent workloads on disjoint hardware subsets.
NVIDIA CUDA Green Contexts~\cite{GreenContexts} are lightweight execution contexts that allow developers to provision subsets of GPU resources, including Streaming Multiprocessors (SMs), and target those resources through CUDA streams and kernel launches.
HEL uses Green Contexts to reserve SMs for its runtime service.
AMD's CPX mode instead exposes compute partitions as separate logical GPUs~\cite{AMD:2025:Partitioning}.
This spatial isolation~\cite{spatialPartitioning} lets task scheduling and dependency management proceed in parallel without interrupting application execution.

\paragraph{Scheduling and Bounded Resources}
Unlike single-kernel GPU task schedulers, HEL must manage overlapping request lifetimes across application kernel launches within a persistent GPU runtime.
Tzeng et al.~\cite{Tzeng:2010:TMI,Tzeng:2012:GTP} study stealing and donation for irregular work distribution, and dependency counters for readiness.
HEL also tracks dependency readiness, but its evaluated task-spawning configuration (Section~\ref{sec:implementation:modes}) reserves separate worker groups for application-submitted and service-spawned tasks without stealing or donation.
For bounded storage, Kokkos Task DAG supports cooperative respawning~\cite{Edwards:2017:KTD}, while Orr et al.\ propose hardware
  wavefront yielding~\cite{Orr:2014:FTA}.
HEL instead uses run-to-completion task invocations and reserves storage for service-spawned tasks per \emph{lineage}---each admitted application-submitted task (the \emph{root}) and its descendants---without hardware yielding.
Unlike the isolated phase/depth record budgets discussed for DirectX Work Graphs~\cite{Microsoft:2026:WorkGraphs}, HEL assigns these reservations by ancestry. 
This prevents resource deadlock caused by unrelated roots exhausting the slots an active task needs to spawn follow-up work and complete (Section~\ref{sec:design:lineages}).

%% file: rs/design.tex
\section{Design Goals, Decisions, and Programming Model}%
\label{sec:design}

\input{rs/task_pseudocode.tex}

Our goal is a clean task-based GPU runtime abstraction that hides scheduling internals, guided by two principles: (1)~implementing the entire control plane on the GPU, and (2)~implementing runtime functionality separably from application functionality.
HEL provides persistent runtime services in software on existing GPUs using a reserved SM partition.
Given current hardware and programming-model limits, reserving fixed resources will not beat traditional CUDA on simple, compute-bound workloads that scale across the entire GPU\@.
While Section~\ref{sec:results} identifies regimes where HEL achieves speedups, we frame this work primarily as a systems exploration of both principles.

Implemented as a persistent service under bounded resources, HEL executes many concurrent, explicitly dependent tasks in parallel with the user application, prioritizing steady-state task throughput and efficiency over single-task latency.
Rather than requiring custom, one-off integration via \texttt{\#include} inside kernels, we factor common runtime functionality into a standalone \emph{service} with a standardized interface.
Services run independently of application code, persist to maintain active state beyond individual kernels, and can be implemented once in a modular framework for reuse across diverse applications.

Separating application and runtime enables concurrent, asynchronous execution on disjoint partitions of the same GPU, improving utilization as GPUs grow too large for even medium-sized workloads to saturate.\footnote{Even ML workloads frequently underutilize GPUs: Guardian~\cite{Pavlidakis:2024:GSG} notes partial utilization, MuxFlow~\cite{Liu:2024:MEG} reports 26\% utilization in ByteDance production clusters, and Elvinger et al.~\cite{Elvinger:2025:UGR} cite Microsoft's $<$10\% compute utilization during Llama3-8B's memory-bound decoding~\cite{Kamath:2025:PAS}.\label{footnote:underutilization}}
HEL offsets fixed orchestration overhead relative to monolithic baselines by (i)~amortizing costs across calls as a two-stage pipeline (application posting + service processing) and (ii)~hiding latency by overlapping service-executed work (orchestration + task execution) with inline application compute, as validated in Section~\ref{sec:results}.

\subsection{Major Design Decisions}

\subsubsection{Runtime Functionality via Spatial Partitioning}
Four strategies can provide user-specified runtime functionality on GPUs:
\paragraph{Direct \#include} Inlining runtime code (e.g., an allocator or compressor) into a monolithic kernel enables compiler optimization, but sacrifices concurrency and modularity.
\paragraph{Uberkernels}
Combining application and runtime kernels via a top-level \texttt{switch} and custom scheduler imposes a restrictive programming structure and tightly couples their execution.
The combined code can increase per-thread register requirements and reduce occupancy.
Separate kernels allow each role to have its own register footprint.
\paragraph{Temporal partitioning}
Alternating full-GPU timeslices between application and runtime kernels prevents concurrent execution and raises interaction latency, even with preemption~\cite{isaacGelado_preemption,multitaskingSurvey}.
\paragraph{Spatial partitioning (HEL)}
Running persistent, stateful runtime services concurrently with the application on a dedicated GPU partition fulfills our functionality, concurrency, and modularity goals, persisting throughout and beyond application execution.

\subsubsection{Application-Runtime Communication}
HEL provides a uniform request interface for modular service integration.
Application threads call services by submitting task requests specifying an operation and its parameters.
Task handles expose completion and optional results.
The interface supports synchronous and asynchronous calls, batching, and automatic retries, while HEL manages scheduling and dependency tracking.

\subsubsection{Bounded Per-Lineage Resource Isolation}
\label{sec:design:lineages}
A task executing in HEL may need to allocate storage for further tasks before returning.
If unrelated requests exhaust a shared pool, allocation can block the tasks whose completion would release storage, creating a resource-deadlock risk.
In task-spawning mode (Section~\ref{sec:implementation:modes}), before executing an application-submitted root, HEL reserves fixed-capacity queues and pools shared by its descendants.
The root and its descendants form a \emph{lineage}.
Slots are reused as tasks and handles are reclaimed, so capacity limits simultaneously live work rather than total tasks generated.
Ownership follows the lineage's lifetime across application launches, preventing unrelated roots from consuming its reserved slots.
Separate worker groups for application-submitted and service-spawned tasks prevent admission waits from occupying all workers needed by descendants (Sections~\ref{sec:implementation:workers} and~\ref{sec:implementation:admission}).

%% file: rs/task_pseudocode.tex
\definecolor{helBaseBg}{HTML}{F8FAFC}
\definecolor{helBaseHdr}{HTML}{E2E8F0}
\definecolor{helBaseBdr}{HTML}{94A3B8}
\definecolor{helBaseTxt}{HTML}{1E293B}
\definecolor{helModBg}{HTML}{F6FEF9}
\definecolor{helModHdr}{HTML}{DCFCE7}
\definecolor{helModBdr}{HTML}{5B9A72}
\definecolor{helModTxt}{HTML}{14532D}
\definecolor{helCodeFg}{HTML}{0F172A}
\definecolor{helKwBlue}{HTML}{1D4ED8}
\definecolor{helKwTeal}{HTML}{0F766E}
\definecolor{helComment}{HTML}{64748B}
\definecolor{helLineNo}{HTML}{94A3B8}
\providecommand{\helcardhdr}[4]{%
  {\setlength{\fboxsep}{2.1pt}\setlength{\fboxrule}{0.55pt}%
   \noindent\fcolorbox{#1}{#2}{%
     \parbox{\dimexpr\linewidth-2\fboxsep-2\fboxrule\relax}{%
       \strut\fontfamily{lmss}\fontsize{6.6}{7.8}\selectfont\color{#3}#4%
     }%
   }}\par\nointerlineskip\vspace{-1.35pt}%
}
\lstdefinestyle{helbase}{
  language=C++,
  basicstyle=\fontfamily{lmtt}\fontsize{6.3}{7.2}\selectfont\color{helCodeFg},
  keywordstyle=\bfseries\color{helKwBlue},
  commentstyle=\itshape\color{helComment},
  emph={HEL,async_request,call,get,coroutine,co_return,co_yield,co_await,spawn,kernel,device,host,procedure},
  emphstyle=\bfseries\color{helKwTeal},
  numbers=left,
  numberstyle=\fontfamily{lmss}\fontsize{5.3}{7.2}\selectfont\color{helLineNo},
  numbersep=4.5pt,
  framerule=0.55pt,
  framesep=2.4pt,
  xleftmargin=14.8pt,
  framexleftmargin=11.85pt,
  xrightmargin=2.95pt,
  framexrightmargin=0pt,
  aboveskip=0pt,
  belowskip=0pt,
  frame=single,
  columns=fullflexible,
  keepspaces=true,
  showstringspaces=false
}
\lstdefinestyle{helleft}{style=helbase,backgroundcolor=\color{helBaseBg},rulecolor=\color{helBaseBdr}}
\lstdefinestyle{helright}{style=helbase,backgroundcolor=\color{helModBg},rulecolor=\color{helModBdr}}

\begin{figure*}[!t]
    \centering
    \begin{minipage}[t]{0.488\linewidth}
        \helcardhdr{helBaseBdr}{helBaseHdr}{helBaseTxt}{\textbf{Coroutine Decomposition} \normalfont\fontfamily{lmss}\fontsize{6.1}{7.8}\selectfont\color{helComment}(Requires Compiler Lowering)}
\begin{lstlisting}[style=helleft]
coroutine Producer(Data* buf, int step) {
  co_return produce(buf, step);      // Suspend; yield item
}
coroutine Consumer(Data* buf, int step) {
  for (; step < K; ++step) {
    auto item = co_await Producer(buf, step);
    consume(buf, item);              // Resume when ready
  }
}
kernel App_Main(Data* buf) {
  spawn Consumer(buf, 0);            // Launch deferred chain
  inline_work(buf);                  // Overlaps deferred work
}
\end{lstlisting}
        \captionsetup{type=algorithm,skip=2.5pt}
        \captionof{algorithm}{Coroutine-style decomposition.}%
        \label{algo:task_pseudocode_coroutine}
    \end{minipage}%
    \hfill
    \begin{minipage}[t]{0.488\linewidth}
        \helcardhdr{helModBdr}{helModHdr}{helModTxt}{\textbf{HEL Task API} \normalfont\fontfamily{lmss}\fontsize{6.1}{7.8}\selectfont\color{helModTxt!75}(Library-Only, Persistent GPU Service)}
\begin{lstlisting}[style=helright]
device void submit_step(Data* buf, int step) {
  auto producer = [buf, step]() { return produce(buf, step); };
  auto consumer = [buf, step](auto item) {
    consume(buf, item);
    if (step + 1 < K) submit_step(buf, step + 1); // SMEM chain
  };
  auto prod_h = HEL.async_request(producer);
  HEL.async_request(prod_h, consumer); // Await; pass item
}
kernel App_Main(Data* buf) {
  submit_step(buf, 0);               // Non-blocking DRAM post
  inline_work(buf);                  // Overlaps HEL service
}
\end{lstlisting}
        \captionsetup{type=algorithm,skip=2.5pt}
        \captionof{algorithm}{Equivalent HEL task API\@.}%
        \label{algo:task_pseudocode_hel}
    \end{minipage}
\end{figure*}

%% file: rs/programming_model.tex
\subsection{HEL Programming Model}%
\label{sec:programming-model}

Service developers implement services as HEL-managed tasks that can execute
concurrently with application work.
Tasks are expressed as C++ device lambdas with captured context (inputs/outputs), optional results, and explicit dependencies, executed at thread, warp, or block granularity.

HEL targets workloads poorly matched to monolithic kernels: explicit dependent tasks, complex divergent control flow, small batches that underutilize the GPU, latency-tolerant work that overlaps independent application computation, and task groups that amortize orchestration across invocations.

CUDA Dynamic Parallelism considers a parent grid complete only after its child grids complete~\cite{NVIDIA:2025:CUDA}.
Atos schedules application work through persistent or non-persistent task kernels.
HEL instead lets application kernels call services without adopting the runtime's worker loop, while submitted work and runtime state can outlive the submitting kernel.

Viewed another way (Algorithms~\ref{algo:task_pseudocode_coroutine}--\ref{algo:task_pseudocode_hel}), HEL provides coroutine-style decomposition without compiler support, mirroring compiler coroutine lowering by separating deferred, dependency-managed tasks from inline control.

Programmers call services through HEL's task-request API, receiving (1)~a task \emph{handle} (a completion promise optionally holding a result) for asynchronous requests, or (2)~the result directly via a synchronous, blocking \texttt{get()}.
Asynchronous requests do not wait for task completion. Submission can fail when queues or pools lack capacity, and a retry API repeats attempts up to a count or until success.
HEL supports explicit task dependencies. 
The evaluated prototype uses one prerequisite and one \emph{awaiting} dependent per handle. 

%% file: rs/implementation.tex
\section{Implementation}%
\label{sec:implementation}

\subsection{Service and Worker Architecture}

HEL implements its persistent service as a persistent CUDA kernel on SMs reserved through Green Contexts (Figure~\ref{fig:hel_overview}).
Application and service communicate through task queues and task handles in shared device DRAM.
Service workers poll queues for ready tasks and publish completion status and optional results through task handles.
\emph{Application-submitted tasks} originate in application kernels.
\emph{Service-spawned tasks} are created by tasks executing within the service.
Both execute on HEL's service workers.
Scheduling structures for service-spawned tasks use on-chip shared memory (SMEM), with HEL-provisioned task-local SMEM scratch for warp or block tasks.

\subsubsection{Service Execution Modes}
\label{sec:implementation:modes}
\emph{Task-spawning mode} allows tasks executing in HEL to spawn further tasks within the same lineage, reserving workers and SMEM capacity for service-spawned work.
Combined with explicit dependencies, this supports coroutine-style continuations.
\emph{Leaf-only mode} disables service-side task spawning and assigns all workers and SMEM budget to application-submitted tasks.

\subsubsection{Worker Partitioning and Dispatch}
\label{sec:implementation:workers}
Task granularity is selected at build time.
We describe the evaluated warp-task configuration.
Task-spawning mode partitions each 32-warp service block between the two task categories (16 warps each by default).
Leaf-only mode assigns all 32 to application-submitted tasks.
Workers drain their warp-ID-assigned queues in turn until application submissions finish and both queue sets empty.
These roles govern queue consumption, while a completing worker may directly execute a newly ready dependent task.

\begin{figure}[!t]
\centering
\input{rs/fig1_hel_architecture_v3.tex}
\caption{HEL architecture, memory hierarchy, and task lifecycle: (1)~submission, (2)~lineage admission and execution, and (3)~completion, dependency resolution, and garbage collection (GC), returning reclaimed storage to HEL's pools.}
\label{fig:hel_overview}
\end{figure}

\subsection{Bounded Runtime Data Structures}
HEL bounds memory usage with fixed-capacity, preallocated multi-producer, multi-consumer (MPMC) queues and pools.
All \emph{tasks}---C++ device lambdas copied via \texttt{memcpy} into preallocated task buffers---reside in DRAM and are fetched by pointer only when executed, keeping on-chip scheduling structures small.
\emph{Handles} reside in DRAM and store a reference counter, dependencies, completion state, and optional results.
\emph{Lineage states} maintain the queues, pools, and counters shared by a root's service-spawned descendants.

\emph{Ready tasks} use multiple bounded MPMC FIFO queues implemented as circular buffers (configurable capacity, default: 16 slots).
Queue control structures reside in DRAM for application-submitted tasks and in SMEM for service-spawned tasks, while task storage remains in DRAM.
FIFO order holds within each queue for a given producer-to-queue mapping, not across queues or application calls.
\emph{Waiting tasks} become runnable out of arrival order as their dependencies complete, so HEL tracks them across multiple unordered MPMC pools to remove satisfied entries without rewriting order.
Multiple MPMC \emph{handle pools} reduce contention, with control structures in DRAM for application-submitted tasks and in SMEM per lineage for service-spawned tasks.
MPMC \emph{state pools} allocate lineage states in DRAM or SMEM, depending on configuration.

\subsection{Task Lifecycle and Resource Admission}
Tasks progress through submission, lineage admission and execution, and completion (Figure~\ref{fig:hel_overview}, steps~1--3).

\subsubsection{Task Submission}
The calling thread uses HEL's request API to acquire preallocated task and handle slots and initialize the task body and execution-context metadata.
The API inserts ready tasks into ready queues and tasks with unresolved dependencies into waiting pools.

\subsubsection{Lineage Admission and Execution}
\label{sec:implementation:admission}
In task-spawning mode, a worker acquires a lineage state before executing an application-submitted task, waiting if none is available.
Once started, each task invocation runs to completion.
Queue and pool capacities bound simultaneously live tasks and handles, not total tasks generated.
Reclaiming earlier tasks and handles enables slot reuse, while retained ancestor state and pending branches increase live-storage demand.
Allocation retries for service-spawned tasks do not yield the requesting worker, so workloads must keep their per-lineage live-storage demand within the configured capacities.

\subsubsection{Completion, Dependency Resolution, and GC}
Task results and completion status are published through handles, removing newly satisfied tasks from waiting pools so the completing worker may directly execute an awaiting dependent task (or enqueue ready tasks into ready queues).
HEL coordinates dependent-task submission with prerequisite completion so that a dependent task is not missed when its prerequisite finishes concurrently.
If the prerequisite has already completed, the dependent task is submitted directly as ready.
Task buffers and result handles have distinct lifetimes.
Post-execution garbage collection releases the completed task's own handle reference and, if present, its reference to its prerequisite's handle. 
Handles are reclaimed when their reference counts reach zero, and a lineage state is reclaimed only when both its outstanding-task and active-user counts reach zero, keeping storage bounded by the configured pool capacities.

%% file: rs/fig1_hel_architecture_v3.tex
\definecolor{helGpuBg}{HTML}{F8FAFC}
\definecolor{helGpuBorder}{HTML}{1E293B}
\definecolor{helAppBg}{HTML}{EFF6FF}
\definecolor{helAppBorder}{HTML}{2563EB}
\definecolor{helSvcBg}{HTML}{F0FDF4}
\definecolor{helSvcBorder}{HTML}{16A34A}
\definecolor{helSmemBg}{HTML}{FFFBEB}
\definecolor{helSmemBorder}{HTML}{D97706}
\definecolor{helDramBg}{HTML}{E2E8F0}
\definecolor{helDramHeader}{HTML}{334155}
\definecolor{helBoxFill}{HTML}{FFFFFF}
\definecolor{helFlowBlue}{HTML}{1D4ED8}
\definecolor{helFlowGreen}{HTML}{15803D}
\definecolor{helFlowTeal}{HTML}{0F766E}
\definecolor{helFlowAmber}{HTML}{B45309}
\definecolor{helFlowPurple}{HTML}{6D28D9}
\resizebox{\linewidth}{!}{%
\renewcommand{\sfdefault}{lmss}%
\renewcommand{\ttdefault}{lmtt}%
\begin{tikzpicture}[
    x=1cm, y=1cm,
    >=Latex,
    every node/.style={font=\sffamily},
    lbl/.style={font=\sffamily\fontsize{8.6}{10.0}\selectfont, align=center},
    sublbl/.style={font=\sffamily\fontsize{7.5}{8.8}\selectfont, align=center},
    tinyLbl/.style={font=\sffamily\fontsize{6.9}{8.1}\selectfont, align=center},
    microLbl/.style={font=\sffamily\fontsize{6.35}{7.4}\selectfont, align=center},
    hdr/.style={font=\sffamily\fontsize{9.8}{11.2}\selectfont\bfseries, align=center},
    subhdr/.style={font=\sffamily\fontsize{8.4}{9.7}\selectfont\bfseries, align=center},
    compBox/.style={draw=#1, fill=helBoxFill, rounded corners=2.5pt, line width=0.7pt},
    stackBack/.style={draw=#1!70, fill=helBoxFill!92!#1, rounded corners=2.5pt, line width=0.55pt},
    partBox/.style n args={2}{draw=#1, fill=#2, rounded corners=3.5pt, line width=0.85pt},
    arrSub/.style={->, line width=0.95pt, draw=helFlowBlue},
    arrRet/.style={->, densely dashed, line width=0.85pt, draw=helFlowBlue},
    arrFetch/.style={->, densely dashed, line width=0.85pt, draw=helFlowTeal},
    arrPop/.style={->, line width=0.95pt, draw=helFlowGreen},
    arrGC/.style={->, line width=0.95pt, draw=helFlowPurple},
    arrSmem/.style={->, line width=0.88pt, draw=helFlowAmber}
]

\filldraw[fill=helGpuBg, draw=helGpuBorder, rounded corners=5pt, line width=1.1pt]
    (0,0) rectangle (16.2,9.65);
\fill[helGpuBorder] (0,9.00) [rounded corners=4pt] rectangle (16.2,9.65);
\fill[helGpuBorder] (0,9.00) rectangle (16.2,9.30);
\node[font=\sffamily\fontsize{10.0}{11.4}\selectfont\bfseries, text=white] at (8.1,9.32)
    {GPU --- Shared Virtual Address Space (NVIDIA CUDA Green Contexts)};

\filldraw[partBox={helAppBorder}{helAppBg}] (0.22,4.34) rectangle (5.12,8.80);
\node[hdr, text=helAppBorder!85!black] at (2.67,8.48) {APP Partition};
\node[tinyLbl, text=helGpuBorder] at (2.67,8.11) {Green Ctx 1 ($N_{\text{app}}$ SMs, Work Queue)};

\filldraw[compBox=helAppBorder!80!black] (0.40,4.52) rectangle (4.94,7.82);
\node[subhdr, text=helGpuBorder] at (2.67,7.53) {Application Kernel(s)};

\filldraw[draw=helGpuBorder!60, fill=helAppBg!55, rounded corners=2pt, line width=0.55pt]
    (0.58,6.30) rectangle (4.76,7.24);
\node[lbl] at (2.67,6.94) {\textbf{Inline Application Work}};
\node[sublbl, text=helGpuBorder] at (2.67,6.55) {Compute Plane (concurrent)};

\filldraw[draw=helAppBorder, fill=white, rounded corners=2pt, line width=0.6pt]
    (0.58,4.70) rectangle (4.76,6.10);
\node[lbl, text=helFlowBlue] at (2.67,5.80) {\textbf{HEL Request API}};
\node[sublbl] at (2.67,5.38) {\texttt{async\_request(deps, lambda)}};
\node[tinyLbl, text=helGpuBorder] at (2.67,4.96) {sync \texttt{get()} / automatic retry};

\filldraw[partBox={helSvcBorder}{helSvcBg}] (5.34,4.34) rectangle (15.98,8.80);
\node[hdr, text=helSvcBorder!80!black] at (10.66,8.48) {SERVICE Partition};
\node[tinyLbl, text=helGpuBorder] at (10.66,8.11) {Green Ctx 2 ($N_{\text{svc}}$ SMs, Dedicated Work Queue)};

\filldraw[compBox=helSvcBorder!85!black] (5.52,4.52) rectangle (10.52,7.82);
\node[subhdr, text=helGpuBorder] at (8.02,7.53) {Persistent Service Kernel};

\filldraw[draw=helFlowAmber!85!black, fill=helSmemBg!45, rounded corners=2pt, line width=0.6pt]
    (5.70,6.00) rectangle (10.34,7.28);
\node[lbl, text=helFlowAmber!90!black] at (8.02,6.95) {\textbf{Workers for}\\\textbf{Service-Spawned Tasks}};
\node[tinyLbl, text=helGpuBorder] at (8.02,6.49) {16 warps/block default (0 in leaf-only)};
\node[microLbl, text=helGpuBorder] at (8.02,6.16) {Direct continuation or SMEM queue run};

\filldraw[draw=helSvcBorder!90!black, fill=white, rounded corners=2pt, line width=0.6pt]
    (5.70,4.68) rectangle (10.34,5.94);
\node[lbl, text=helFlowGreen!90!black] at (8.02,5.61) {\textbf{Workers for}\\\textbf{Application-Submitted Tasks}};
\node[tinyLbl, text=helGpuBorder] at (8.02,5.16) {16 warps/block default (32 in leaf-only)};
\node[microLbl, text=helGpuBorder] at (8.02,4.85) {Reserve lineage state $\rightarrow$ run to completion};

\filldraw[partBox={helSmemBorder}{helSmemBg}] (11.64,4.52) rectangle (15.80,7.82);
\node[subhdr, text=helFlowAmber] at (13.72,7.53) {Per-Lineage State (SMEM)};
\node[microLbl, text=helGpuBorder] at (13.72,7.20) {Reserved on admission (+ scratch)};

\filldraw[stackBack=helSmemBorder] (11.90,6.01) rectangle (15.64,7.01);
\filldraw[compBox=helSmemBorder] (11.82,5.93) rectangle (15.56,6.93);
\node[tinyLbl] at (13.69,6.68) {\textbf{Ready Queues}\\\textbf{(Service-Spawned Tasks)}};
\node[microLbl, text=helGpuBorder] at (13.69,6.31) {MPMC FIFO, 16 slots per queue};
\foreach \i in {0,1,2,3} {
    \draw[fill=helSmemBg, draw=helSmemBorder, line width=0.55pt] ({12.42+\i*0.52},6.00) rectangle ({12.70+\i*0.52},6.15);
}
\node[sublbl] at (14.84,6.07) {$\cdots$};

\filldraw[stackBack=helSmemBorder] (11.90,4.76) rectangle (15.64,5.64);
\filldraw[compBox=helSmemBorder] (11.82,4.68) rectangle (15.56,5.56);
\node[tinyLbl] at (13.69,5.34) {\textbf{WAITING \& Handle Pools}};
\node[microLbl, text=helGpuBorder] at (13.69,5.08) {Unordered $O(1)$ pools + counters};
\foreach \i in {0,1,2,3} {
    \draw[fill=helSmemBg, draw=helSmemBorder, line width=0.55pt] ({12.56+\i*0.52},4.85) circle (0.095);
}
\node[sublbl] at (14.84,4.85) {$\cdots$};

\draw[arrSmem] (13.69,5.64) -- node[right, microLbl, text=helFlowAmber, xshift=2pt]{\texttt{deps}$=0$} (13.69,5.93);
\draw[arrSmem] (11.64,6.58) -- node[above, tinyLbl, text=helFlowAmber]{pop ptr} (10.52,6.58);
\draw[arrSmem] (10.52,5.22) -- node[above, microLbl, text=helFlowAmber]{reserve /} node[below, microLbl, text=helFlowAmber]{spawn} (11.64,5.22);

\filldraw[partBox={helDramHeader}{helDramBg}] (0.22,0.20) rectangle (15.98,2.94);
\fill[helDramHeader] (0.22,0.20) [rounded corners=3pt] rectangle (15.98,0.66);
\fill[helDramHeader] (0.22,0.44) rectangle (15.98,0.66);
\node[font=\sffamily\fontsize{8.7}{10.2}\selectfont\bfseries, text=white] at (8.1,0.43)
    {Global Device Memory (DRAM --- Shared Across Green Contexts)};

\filldraw[stackBack=helDramHeader] (0.48,0.88) rectangle (5.16,2.80);
\filldraw[compBox=helDramHeader] (0.40,0.80) rectangle (5.08,2.72);
\node[subhdr, text=helGpuBorder] at (2.74,2.45) {Task Buffers \& Multi-Pools};
\node[tinyLbl] at (2.74,2.06) {\textbf{Task Buffers:} all \texttt{memcpy} lambdas};
\node[microLbl, text=helGpuBorder] at (2.74,1.67) {\textbf{Handle Pools:} \texttt{refcount}, \texttt{deps}, \texttt{result}};
\node[microLbl, text=helGpuBorder] at (2.74,1.34) {\textbf{State Pools:} lineage states ($O(1)$ MPMC)};
\node[microLbl, text=helFlowTeal] at (2.74,0.99) {(lambdas fetched by pointer only on run)};

\filldraw[stackBack=helDramHeader] (6.54,0.88) rectangle (10.62,2.80);
\filldraw[compBox=helDramHeader] (6.46,0.80) rectangle (10.54,2.72);
\node[subhdr, text=helGpuBorder] at (8.50,2.45) {Unordered WAITING Pools};
\node[tinyLbl, text=helGpuBorder] at (8.50,2.06) {Multiple MPMC pools (\texttt{deps} $> 0$)};
\node[microLbl, text=helGpuBorder] at (8.50,1.70) {Out-of-order $O(1)$ removal};
\foreach \i in {0,1,2,3} {
    \draw[fill=helDramBg!65, draw=helDramHeader, line width=0.55pt] ({7.36+\i*0.56},1.18) circle (0.13);
}
\node[sublbl] at (9.78,1.18) {$\cdots$};

\filldraw[stackBack=helDramHeader] (11.98,0.88) rectangle (15.82,2.80);
\filldraw[compBox=helDramHeader] (11.90,0.80) rectangle (15.74,2.72);
\node[subhdr, text=helGpuBorder] at (13.82,2.24) {Ready Queues for\\Application-Submitted\\Tasks};
\node[tinyLbl, text=helGpuBorder] at (13.82,1.65) {MPMC FIFO, 16 slots per queue};
\node[microLbl, text=helGpuBorder] at (13.82,1.34) {Submitted (\texttt{deps}$=0$) or promoted};
\foreach \i in {0,1,2,3,4} {
    \draw[fill=helDramBg!65, draw=helDramHeader, line width=0.55pt] ({12.38+\i*0.50},0.96) rectangle ({12.66+\i*0.50},1.13);
}
\node[sublbl] at (15.18,1.05) {$\cdots$};

\draw[->, line width=0.85pt, draw=helFlowBlue] (5.16,2.20) -- (6.46,2.20);
\node[microLbl, text=helFlowBlue] at (5.81,2.43) {if \texttt{deps}$>0$};
\node[microLbl, text=helFlowBlue] at (5.81,1.97) {(else ready)};

\draw[->, line width=0.85pt, draw=helFlowPurple] (6.46,1.22) -- (5.16,1.22);
\node[microLbl, text=helFlowPurple] at (5.81,1.45) {GC free};
\node[microLbl, text=helFlowPurple] at (5.81,0.98) {(\texttt{refs}$=0$)};

\draw[->, line width=0.95pt, draw=helFlowPurple] (10.62,1.74) -- (11.90,1.74);
\node[microLbl, text=helFlowPurple] at (11.26,2.18) {\textbf{promote}};
\node[microLbl, text=helFlowPurple] at (11.26,1.94) {or direct run};
\node[microLbl, text=helFlowPurple] at (11.26,1.48) {(\texttt{deps}$=0$)};

\draw[arrRet] (1.18,2.80) -- (1.18,4.52);
\node[tinyLbl, text=helFlowBlue, anchor=east] at (1.10,3.66) {\texttt{handle}\\$/$ \texttt{get()}};

\draw[arrSub] (3.05,4.52) -- (3.05,2.80);
\node[sublbl, text=helFlowBlue, fill=helGpuBg, inner sep=1.5pt] at (3.05,3.82) {\textbf{(1) submit}};
\node[microLbl, text=helFlowBlue, fill=helGpuBg, inner sep=1pt] at (3.05,3.40) {\texttt{memcpy} + handle};

\draw[arrFetch] (4.72,2.80) -- (4.72,3.62) -| (6.32,4.52);
\node[microLbl, text=helFlowTeal, fill=helGpuBg, inner sep=1.2pt] at (5.52,3.62) {fetch \texttt{lambda} on run};

\draw[arrGC] (8.15,4.52) -- (8.15,2.80);
\node[sublbl, text=helFlowPurple, fill=helGpuBg, inner sep=1.5pt] at (8.15,3.66) {\textbf{(3) complete \& GC}};

\draw[arrPop] (13.30,2.80) -- (13.30,3.62) -| (10.05,4.52);
\node[sublbl, text=helFlowGreen!90!black, fill=helGpuBg, inner sep=1.5pt] at (11.70,3.62) {\textbf{(2) admit \& run}};

\end{tikzpicture}%
}

%% file: rs/tests.tex
\section{Evaluation}%
\label{sec:evaluation}

Experiments use an AMD EPYC 7402 CPU, an NVIDIA A100-PCIE-40GB GPU (108~SMs), and \texttt{nvcc}~13.0.48, following one runtime and one application warmup run.
Modular applications share the GPU with HEL, while monolithic baselines use the full GPU.
All HEL experiments use warp-level tasks.

\input{rs/eval_benchmarks.tex}

\subsection{Producer-Consumer Microbenchmark}
\label{sec:evaluation:producer-consumer}
The monolithic baseline (Algorithm~\ref{algo:pc}, left) uses a host-controlled loop to launch separate producer and consumer kernels across the entire GPU on each iteration.
The host-side \texttt{HEL.call} wrapper launches an application kernel running \texttt{PC\_mod} with a pointer to HEL's runtime (Algorithm~\ref{algo:pc}, right).
The application calls the service via \texttt{rs->async\_request} with \texttt{producer\_task} and \texttt{consumer\_task} lambdas, passing the producer's handle as the consumer's dependency.
The consumer processes the producer's result and, before returning, calls \texttt{PC\_mod} if iterations remain, spawning the next pair through two \texttt{async\_request} calls.
Both versions use $R$ application threads, one per independent chain of $N$ producer-consumer pairs ($\mathrm{producer}_1 \to \mathrm{consumer}_1 \to \mathrm{producer}_2 \to \mathrm{consumer}_2 \to \cdots$).

Eliminating global Producer-Consumer synchronization, HEL schedules ready tasks dynamically, transforming the $N$-iteration loop into service-side task generation that emulates compiler coroutine lowering (which native GPU coroutines could automate).
The application kernel can finish without waiting for its service tasks, allowing subsequent launches to overlap ongoing service work.
Below, ``load'' ($R \times N \times \text{context-buffer size}$) serves as a proxy for total work.

\subsection{Host-Remote Procedure Call (Host-RPC) Microbenchmark}

Host-RPC models concurrent host-side work to evaluate how HEL hides host latency behind GPU execution as per-task host wait time varies.
In the monolithic baseline (Algorithm~\ref{algo:host_rpc}, left), threads spin-wait on a host-completion flag before executing work, whereas in HEL (Algorithm~\ref{algo:host_rpc}, right), runtime tasks (\texttt{host\_wait\_task}) wait on the flag while the application concurrently executes inline work.

\input{rs/results}

%% file: rs/eval_benchmarks.tex
\definecolor{helBaseBg}{HTML}{F8FAFC}
\definecolor{helBaseHdr}{HTML}{E2E8F0}
\definecolor{helBaseBdr}{HTML}{94A3B8}
\definecolor{helBaseTxt}{HTML}{1E293B}
\definecolor{helModBg}{HTML}{F6FEF9}
\definecolor{helModHdr}{HTML}{DCFCE7}
\definecolor{helModBdr}{HTML}{5B9A72}
\definecolor{helModTxt}{HTML}{14532D}
\definecolor{helCodeFg}{HTML}{0F172A}
\definecolor{helKwBlue}{HTML}{1D4ED8}
\definecolor{helKwTeal}{HTML}{0F766E}
\definecolor{helComment}{HTML}{64748B}
\definecolor{helLineNo}{HTML}{94A3B8}
\providecommand{\helcardhdr}[4]{%
  {\setlength{\fboxsep}{2.1pt}\setlength{\fboxrule}{0.55pt}%
   \noindent\fcolorbox{#1}{#2}{%
     \parbox{\dimexpr\linewidth-2\fboxsep-2\fboxrule\relax}{%
       \strut\fontfamily{lmss}\fontsize{6.6}{7.8}\selectfont\color{#3}#4%
     }%
   }}\par\nointerlineskip\vspace{-1.35pt}%
}
\lstdefinestyle{helbase}{
  language=C++,
  basicstyle=\fontfamily{lmtt}\fontsize{6.3}{7.2}\selectfont\color{helCodeFg},
  keywordstyle=\bfseries\color{helKwBlue},
  commentstyle=\itshape\color{helComment},
  emph={HEL,async_request,call,get,coroutine,co_return,co_yield,co_await,spawn,kernel,device,host,procedure},
  emphstyle=\bfseries\color{helKwTeal},
  numbers=left,
  numberstyle=\fontfamily{lmss}\fontsize{5.3}{7.2}\selectfont\color{helLineNo},
  numbersep=4.5pt,
  framerule=0.55pt,
  framesep=2.4pt,
  xleftmargin=14.8pt,
  framexleftmargin=11.85pt,
  xrightmargin=2.95pt,
  framexrightmargin=0pt,
  aboveskip=0pt,
  belowskip=0pt,
  frame=single,
  columns=fullflexible,
  keepspaces=true,
  showstringspaces=false
}
\lstdefinestyle{helleft}{style=helbase,backgroundcolor=\color{helBaseBg},rulecolor=\color{helBaseBdr}}
\lstdefinestyle{helright}{style=helbase,backgroundcolor=\color{helModBg},rulecolor=\color{helModBdr}}

\begin{figure*}[!t]
    \captionsetup{type=algorithm,skip=2pt}
    \centering
    \begin{minipage}[t]{0.488\linewidth}
        \helcardhdr{helBaseBdr}{helBaseHdr}{helBaseTxt}{\textbf{Monolithic Baseline} \normalfont\fontfamily{lmss}\fontsize{6.1}{7.8}\selectfont\color{helComment}(Host-Driven Loop, All 108 SMs)}
\begin{lstlisting}[style=helleft]
device int producer() { return 12321; }
device void consumer(int* buf, int v) { *buf = v; }
kernel Producer_kernel(int* buf, int* ctx) {
  write_context(ctx);                // Scales compute/mem
  buf[tid] = producer();
}
kernel Consumer_kernel(int* buf, int* ctx) {
  consumer(&buf[tid], buf[tid] * read_context(ctx));
}
host void PC_mono(int R, int N) {    // 2N host launches
  for (int d = 0; d < N; ++d) {
    Producer_kernel<<<R/T, T>>>(buf, ctx);
    Consumer_kernel<<<R/T, T>>>(buf, ctx);
  }
}
\end{lstlisting}
    \end{minipage}%
    \hfill
    \begin{minipage}[t]{0.488\linewidth}
        \helcardhdr{helModBdr}{helModHdr}{helModTxt}{\textbf{HEL Modular} \normalfont\fontfamily{lmss}\fontsize{6.1}{7.8}\selectfont\color{helModTxt!75}(Disjoint App \& Service Green Contexts)}
\begin{lstlisting}[style=helright]
device void PC_mod(Runtime* rs, int* buf, int* ctx, int step, int N) {
  write_context(ctx);                // Init before capture
  auto producer_task = []() { return producer(); };
  auto consumer_task = [rs, buf, ctx, step, N](int prod) {
    consumer(buf, prod * read_context(ctx));
    if (step + 1 < N) PC_mod(rs, buf, ctx, step + 1, N);
  };
  auto prod_h = rs->async_request(producer_task);
  rs->async_request(prod_h, consumer_task);
}
// rs points to HEL's device runtime; service already active.
HEL.call(PC_mod, R, rs, buf, ctx, 0, N); // N = depth + 1
host void HEL::call(AppFn fn, int R, Args... args) {
  app_kernel<<<B_app, W_app, 0, app_stream>>>(fn, R, args...);
}
\end{lstlisting}
    \end{minipage}
    \captionof{algorithm}{Producer-Consumer microbenchmark: monolithic baseline (left) vs.\ GPU-orchestrated HEL modular pipeline (right).}%
    \label{algo:pc}

    \vspace{3.5pt}

    \begin{minipage}[t]{0.488\linewidth}
        \helcardhdr{helBaseBdr}{helBaseHdr}{helBaseTxt}{\textbf{Monolithic Baseline} \normalfont\fontfamily{lmss}\fontsize{6.1}{7.8}\selectfont\color{helComment}(Serial Host-Flag Wait Then Compute)}
\begin{lstlisting}[style=helleft]
kernel Compute_kernel(int* ctx) {
  inline_compute(ctx);               // Runs after flag set
}
procedure Host_RPC_mono(volatile int* flag, int* ctx) {
  while (*flag == 0) nanosleep();    // Blocks inline compute
  Compute_kernel<<<R/T, T>>>(ctx);
}
\end{lstlisting}
    \end{minipage}%
    \hfill
    \begin{minipage}[t]{0.488\linewidth}
        \helcardhdr{helModBdr}{helModHdr}{helModTxt}{\textbf{HEL Modular} \normalfont\fontfamily{lmss}\fontsize{6.1}{7.8}\selectfont\color{helModTxt!75}(Service Waits While App Computes Inline)}
\begin{lstlisting}[style=helright]
auto host_wait_task = [](volatile int* flag) {
  while (*flag == 0) nanosleep();    // Runs on Service Ctx
};
device void Host_RPC_mod(volatile int* flag, int* ctx) {
  HEL.async_request(host_wait_task, flag);
  inline_compute(ctx);               // Overlaps host wait!
}
\end{lstlisting}
    \end{minipage}
    \captionof{algorithm}{Host-RPC microbenchmark: monolithic baseline (left) vs.\ overlapped HEL modular execution (right).}%
    \label{algo:host_rpc}
\end{figure*}

%% file: rs/results.tex
\begin{figure}[tbp]
  \centering
  \includegraphics[width=\columnwidth]{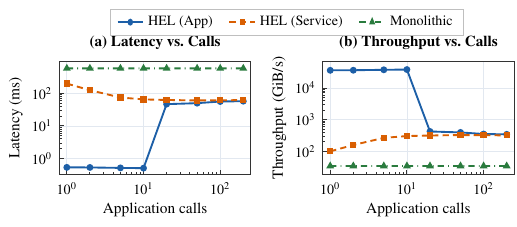}
  \caption{HEL outperforms the baseline in the mixed regime with control-plane pressure when amortized.}
  \label{plots:amort_q1}
\end{figure}

\begin{figure*}[!t]
  \centering
  \begin{subfigure}{\linewidth}
    \centering
    \includegraphics[width=\linewidth]{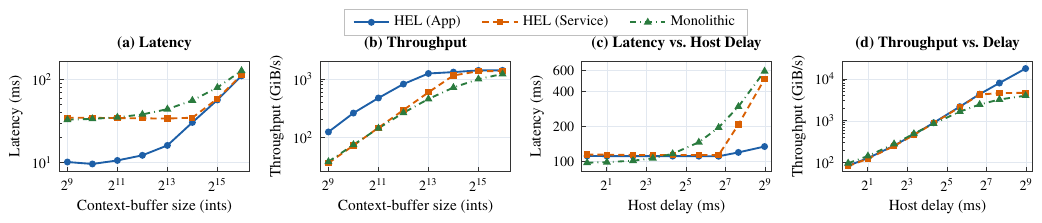}
  \end{subfigure}
  \caption{HEL outperforms the baseline in the latency-hiding regime both when inline compute increases and when host-side delay increases, because the modular execution can overlap waiting time with useful work.}%
  \label{plots:host_rpc_q1}
\end{figure*}

\begin{figure*}[!t]
  \centering
\begin{minipage}[t]{\columnwidth}
    \centering
    \includegraphics[width=\linewidth]{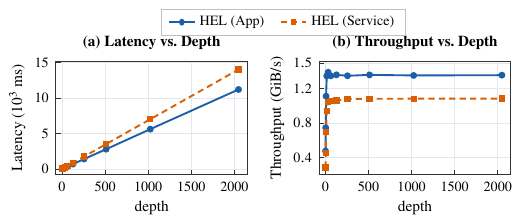}
  \caption{Latency increases as the task-dependency structure deepens, while throughput quickly reaches a plateau.}%
  \label{plots:recdepth_q2}
  \end{minipage}\hfill
  \begin{minipage}[t]{\columnwidth}
    \centering
	\includegraphics[width=\linewidth]{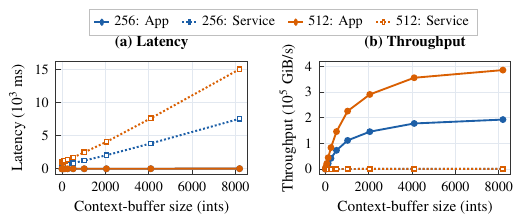}
  \caption{Context-buffer size sweep in the control-heavy regime for two lineage sizes. Latency increases with buffer size, while throughput shows no clear plateau before memory capacity limits the sweep.}%
  \label{plots:context_q2}
  \end{minipage}
\end{figure*}

\subsection{Results}%
\label{sec:results}

HEL's runtime service runs dynamically created tasks with dynamic dependencies and optional asynchronous execution alongside an application, outperforming a monolithic baseline when the GPU is underutilized or latency can be hidden, and tying or trailing when the baseline saturates hardware.
We report three mean per-call latency curves---\emph{monolithic} (end-to-end baseline), \emph{modular} or \emph{app} (HEL application), and \emph{service} (amortized service execution per call)---plus throughput under the same per-application-call normalization.

\paragraph{System Amortization}
\label{sec:results:amortization}
Varying sequential \texttt{PC\_mod} calls (Algorithm~\ref{algo:pc}, right) against one persistent HEL service measures pipeline benefits and fixed-overhead amortization under sustained load.
We define \emph{depth} $d=N-1$ as the number of producer-consumer pairs generated after the initial pair requested by the application.
\emph{Context}-buffer size controls per-task computation and memory footprint.
This experiment uses depth~256 with substantial per-task work (4~kB context buffer)---a control-heavy mixed regime---partitioned via CUDA Green Contexts into \{44, 64\}~SMs for \{application, runtime\} while leaving the GPU unsaturated.
In Figure~\ref{plots:amort_q1}, unpipelined baseline latency remains constant, whereas modular latency rises with call count as queues fill and saturate.
Mean service latency starts high (fixed overhead spread over few calls) and plateaus slightly above modular latency due to request waiting and processing.
Correspondingly, service throughput rises with pipeline fill to a steady-state plateau, while modular throughput starts high, drops under backpressure, and converges to the sustainable service rate---demonstrating how control-plane-heavy workloads benefit from offloading scheduling and dependency management to the persistent service.


\paragraph{Latency Hiding}
Using Host-RPC (Algorithm~\ref{algo:host_rpc}) amortized over 100 calls of 10k requests each (unsaturated baseline), Figure~\ref{plots:host_rpc_q1} characterizes latency hiding along two complementary sweeps.
Sweeping context-buffer size at a fixed 32~ms host delay (Figure~\ref{plots:host_rpc_q1}(a--b)) exposes the overlap boundary: because host delay and baseline compute do not overlap, baseline latency grows roughly linearly with buffer size, whereas modular latency stays nearly constant until application compute exceeds the 32~ms host-delay window.
Sweeping host delay at 256~kB inline compute (Figure~\ref{plots:host_rpc_q1}(c--d)), service latency scales with host delay while baseline throughput saturates below service throughput.
Together, both sweeps show HEL hiding host-wait time as inline compute or host delay grows, benefiting workloads with GPU-initiated communication or externally delayed stages whenever useful inline GPU work can overlap waiting.

All remaining sweeps run Producer-Consumer (Algorithm~\ref{algo:pc}), and all Producer-Consumer microbenchmarks in this section use warp-level tasks in task-spawning mode, where deeper dependencies raise control-plane pressure and larger context buffers raise compute-plane pressure.
Amortization, Host-RPC, depth, and context-buffer size sweeps assign 2 warps per service block to application-submitted tasks.
The \texttt{num\_requests} and SM-partition sweeps instead assign 16 warps to each task category per 1024-thread service block.

\paragraph{Depth of Task Dependencies}
In Figure~\ref{plots:recdepth_q2}, increasing depth raises both per-request work and control-plane pressure.
The horizontal axis reports depth~$d$ (Section~\ref{sec:results:amortization}), with depth~0 executing only the initial pair and depth~1 adding one pair.
Each \texttt{PC\_mod} invocation returns without waiting for its requested pair to complete, and the consumer initiates the next pair if needed.
Thus, each chain executes $N=d+1$ producer and consumer pairs without a growing CUDA call stack or keeping all $2N$ tasks live.
Completed task buffers and unreferenced handles are reclaimed as execution advances, so depth does not count simultaneously allocated tasks or handles.
With warp-level tasks, throughput climbs rapidly up to depth~$\approx 32$ and plateaus, while app and service latency keep rising with lineage length: under a fixed runtime configuration, deeper lineages expose no extra concurrency once ready-work capacity saturates and merely lengthen each request, exposing a control-plane-dominant regime of rising latency at flat throughput.

\paragraph{Lineage Sizes}
Fixing 262k requests at depths 256 and 512, Figure~\ref{plots:context_q2} sweeps context-buffer size to isolate compute cost at large fixed lineages, reaching no clear throughput plateau before memory limits.
Under warp-level submission, doubling depth from 256 to 512 leaves app latency nearly flat, roughly doubles service latency, and holds service throughput nearly constant; app-normalized throughput thus scales almost directly with depth, whereas the service curve better reflects the underlying execution rate.
Deeper lineages are paid almost entirely on the service path while larger context buffers raise compute cost at fixed depth, separating service-side control expansion from buffer-size-driven compute.

\begin{figure*}[!t]
  \centering
	\begin{minipage}[t]{\columnwidth}
    \centering
    \includegraphics[width=\linewidth]{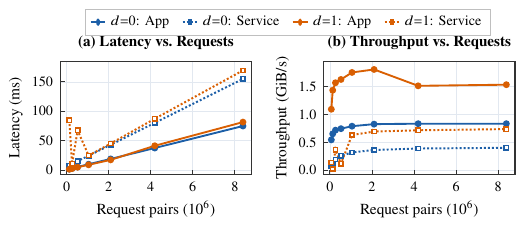}
  \caption{Latency and throughput vs.\ Producer-Consumer task count for depth~0 (minimal runtime) and depth~1 (first-order).}%
  \label{plots:numreq_q4}
\end{minipage}\hfill
 \begin{minipage}[t]{\columnwidth}
 \begin{minipage}[t]{0.485\linewidth}
    \centering
    \includegraphics[width=\linewidth]{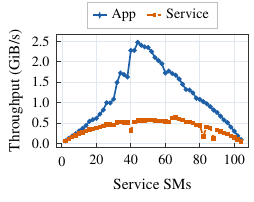}
  \caption{Throughput sensitivity to the SM split between application and HEL.}%
  \label{plots:sm_split_q3}
\end{minipage}\hfill
 \begin{minipage}[t]{0.485\linewidth}
  \centering
  \includegraphics[width=\linewidth]{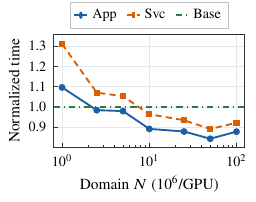}
  \caption{PFPL normalized execution time vs.\ domain size $N$.}%
  \label{fig:size_sweep}
\end{minipage}
\end{minipage}
\end{figure*}

\input{algo/pfpl_compare}

\paragraph{Throughput-Latency Tradeoff under Load}
With one-element context buffers, Figure~\ref{plots:numreq_q4} sweeps independent Producer-Consumer pairs at depth~0 (minimal runtime path) and depth~1 (first service-spawned pair).
These settings execute one and two Producer-Consumer pairs per chain, respectively, with only the latter requiring service-side task spawning.
Throughput rises quickly and plateaus while latency grows roughly linearly under backpressure.
With this balanced worker partition in task-spawning mode, close app-side curves show that service-side task spawning is absorbed primarily on the service path rather than in root-task submission.

\paragraph{Varying the Spatial Partition Size}
For a moderate 1M requests with shallow dependencies (depth~1) and one-element context buffers, Figure~\ref{plots:sm_split_q3} sweeps the GPU SM split.
Throughput rises as service SMs relieve the runtime bottleneck, reaches a broad midrange optimum, and drops once the application partition becomes too small: in this warp-level Producer-Consumer benchmark, app throughput peaks near 44 service SMs while service throughput favors a more service-heavy split, depending on whether application-submitted task posting or service-spawned task execution limits throughput.

\paragraph{Performance Summary}
HEL performs best when its two-stage pipeline sustains steady-state throughput across repeated calls or hides latency via overlap, and worst when a monolithic baseline already saturates the GPU\@.

%% file: algo/pfpl_compare.tex
\definecolor{helBaseBg}{HTML}{F8FAFC}
\definecolor{helBaseHdr}{HTML}{E2E8F0}
\definecolor{helBaseBdr}{HTML}{94A3B8}
\definecolor{helBaseTxt}{HTML}{1E293B}
\definecolor{helModBg}{HTML}{F6FEF9}
\definecolor{helModHdr}{HTML}{DCFCE7}
\definecolor{helModBdr}{HTML}{5B9A72}
\definecolor{helModTxt}{HTML}{14532D}
\definecolor{helCodeFg}{HTML}{0F172A}
\definecolor{helKwBlue}{HTML}{1D4ED8}
\definecolor{helKwTeal}{HTML}{0F766E}
\definecolor{helComment}{HTML}{64748B}
\definecolor{helLineNo}{HTML}{94A3B8}
\providecommand{\helcardhdr}[4]{%
  {\setlength{\fboxsep}{2.1pt}\setlength{\fboxrule}{0.55pt}%
   \noindent\fcolorbox{#1}{#2}{%
     \parbox{\dimexpr\linewidth-2\fboxsep-2\fboxrule\relax}{%
       \strut\fontfamily{lmss}\fontsize{6.6}{7.8}\selectfont\color{#3}#4%
     }%
   }}\par\nointerlineskip\vspace{-1.35pt}%
}
\lstdefinestyle{helbase}{
  language=C++,
  basicstyle=\fontfamily{lmtt}\fontsize{6.3}{7.2}\selectfont\color{helCodeFg},
  keywordstyle=\bfseries\color{helKwBlue},
  commentstyle=\itshape\color{helComment},
  emph={HEL,async_request,call,get,coroutine,co_return,co_yield,co_await,spawn,kernel,device,host,procedure},
  emphstyle=\bfseries\color{helKwTeal},
  numbers=left,
  numberstyle=\fontfamily{lmss}\fontsize{5.3}{7.2}\selectfont\color{helLineNo},
  numbersep=4.5pt,
  framerule=0.55pt,
  framesep=2.4pt,
  xleftmargin=14.8pt,
  framexleftmargin=11.85pt,
  xrightmargin=2.95pt,
  framexrightmargin=0pt,
  aboveskip=0pt,
  belowskip=0pt,
  frame=single,
  columns=fullflexible,
  keepspaces=true,
  showstringspaces=false
}
\lstdefinestyle{helleft}{style=helbase,backgroundcolor=\color{helBaseBg},rulecolor=\color{helBaseBdr}}
\lstdefinestyle{helright}{style=helbase,backgroundcolor=\color{helModBg},rulecolor=\color{helModBdr}}

\begin{figure*}[!t]
    \captionsetup{type=algorithm,skip=2.5pt}
    \centering
    \begin{minipage}[t]{0.488\linewidth}
        \helcardhdr{helBaseBdr}{helBaseHdr}{helBaseTxt}{\textbf{Monolithic Baseline} \normalfont\fontfamily{lmss}\fontsize{6.1}{7.8}\selectfont\color{helComment}(Synchronous Full-GPU Compression)}
\begin{lstlisting}[style=helleft]
// Host blocks stencil at every checkpoint emission:
host void Baseline() {
  for (int t = 1; t <= total_timesteps; ++t) {
    stencil_step<<<108_SMs>>>();
    if (t % emit_interval == 0)
      PFPL_encode<<<108_SMs>>>();
  }
}
\end{lstlisting}
    \end{minipage}%
    \hfill
    \begin{minipage}[t]{0.488\linewidth}
        \helcardhdr{helModBdr}{helModHdr}{helModTxt}{\textbf{HEL Modular} \normalfont\fontfamily{lmss}\fontsize{6.1}{7.8}\selectfont\color{helModTxt!75}(104-SM Stencil Overlaps 4-SM PFPL Service)}
\begin{lstlisting}[style=helright]
auto ckpt = [](int id) { PFPL_compress(id); };
device void HEL_checkpointing() {
  for (int t = 1; t <= total_timesteps; ++t) {
    stencil_step();
    if (t % emit_interval == 0)
      HEL.async_request(ckpt, chunk_id);
  }
}
\end{lstlisting}
    \end{minipage}
    \caption{Checkpoint Compression case study: synchronous baseline (left) vs.\ asynchronous HEL service (right).}
    \label{algo:pfpl_compare}
\end{figure*}

%% file: rs/case_study.tex
\section{Case Study: Checkpoint Compression Service}%
\label{sec:compression}

We evaluate HEL on a realistic compute-bound HPC proxy workload: a heavy 3D stencil with periodic lossy checkpoint compression of the live domain.

\paragraph{Application}
The application runs a 3D 25-point, eighth-order acoustic-wave stencil~\cite{ConvStencil,EBISU} (25 multiplies and 24 adds per cell; four-cell halo per axis) for 1000 timesteps, emitting checkpoints every 100 timesteps in 1024-double chunks compressed with PFPL~\cite{PFPL}, a four-stage lossy floating-point compressor with guaranteed absolute error bound (integer quantization, neighbor-difference encoding, byte transposition, and zero run-length elimination).
Because PFPL operates on independently compressible chunks and includes warp-collective stages, it maps naturally to HEL's warp-granularity service mode.
HEL runs the stencil on 104~SMs while a 4-SM service partition compresses chunks asynchronously, keeping the application-side checkpointing delta small; the monolithic baseline launches synchronous CPU-triggered compression across all 108~SMs at each emission point.
Algorithm~\ref{algo:pfpl_compare} contrasts the synchronous baseline with HEL's asynchronous checkpoint submission.

\paragraph{Results}

Sweeping domain size $N$ from $10^6$ to $10^8$ cells (Figure~\ref{fig:size_sweep}), execution time scales linearly with $N$.
For $N \ge 10^7$, the application path runs at $0.83$--$0.89\times$ baseline time ($1.12$--$1.19\times$ application-path speedup, crossing parity at $2.5 \times 10^6$ cells), whereas the service path runs at $0.89$--$0.96\times$ and crosses parity only near $10^7$ cells---showing that overlap hides service cost before raw service execution beats the baseline.
At $N = 10^6$, fixed runtime overhead remains visible ($0.91\times$ speedup on the application path, $0.76\times$ on the service path).
Although parallel chunk processing resets PFPL's predictor at chunk boundaries and lowers compression ratio, HEL still reduces application-path time for sufficiently large domains: the speedup stems from runtime/application overlap rather than algorithmic improvement.

\paragraph{Structural Advantages Beyond Latency}
Beyond latency, the service abstraction provides structural benefits independent of any single benchmark: decoupling application and compressor code (\emph{modularity}), co-locating multiple services on one partition without inflating kernel complexity (\emph{composability}), preserving state across invocations (\emph{persistence}), enforcing explicit CUDA Green Context resource splits rather than implicit time-multiplexing (\emph{governance}), and deploying new compressors without recompiling applications (\emph{upgradability}).
Like operating systems over bare-metal stacks, these structural advantages motivate HEL even where a hand-tuned monolithic kernel matches or exceeds its raw latency.

%% file: rs/conclusions.tex
\section{Conclusions and Future Work}%
\label{sec:conclusions}

We presented HEL, a modular GPU runtime built on a persistent service that executes dependent, coroutine-style tasks concurrently with inline application work under bounded resources.
HEL differs fundamentally from prior GPU systems across four dimensions:
\begin{description}[leftmargin=0pt, labelsep=0.45em]
\item[Host-Device Decoupling] Unlike CPU-managed runtimes (e.g., StarPU, Legion) that notify the host via interrupts or polling to schedule each subsequent task across PCIe, HEL moves the entire control loop onto the GPU for autonomous task submission and execution.
\item[Spatial vs.\ Temporal] Rather than time-slicing the full GPU across competing kernels, HEL uses NVIDIA CUDA Green Contexts to dedicate fixed SMs to the service, keeping orchestration responsive even when application kernels saturate the remaining hardware.
\item[Independence and Persistence] As a separable entity not compiled into an application's kernel loop, HEL persists across kernel boundaries to manage state, retries, and dependencies on the GPU without host intervention.
\item[Coroutine-Style Suspension] HEL represents explicit suspension points through dependent tasks. 
Once the current task returns, its continuation can remain waiting without occupying a worker, allowing other ready work to execute until the dependency is satisfied. 
\end{description}

Larger SMEM capacity could retain more task and continuation state on chip, while controlled cross-Green-Context access to service-owned SMEM could reduce DRAM traffic.
Compiler support for device coroutines would automate continuation generation but still require storage for suspended state.
These extensions would support persistent software services without replacing them with dedicated runtime hardware.

HEL offers a modular alternative to monolithic kernels that improves performance in pipelined, latency-bound, and underutilized regimes where overlap hides execution cost, while naturally trailing hand-tuned monolithic kernels that already saturate the GPU\@.
Although tightly coupled CPU-GPU architectures (e.g., NVIDIA Grace Hopper, AMD MI300A) reduce discrete-GPU host-launch penalties, ongoing GPU underutilization\textsuperscript{\ref{footnote:underutilization}} makes persistent device services increasingly valuable.
Future multi-node work asks two questions: is network communication a viable HEL service?
Can HEL's abstraction hide from application programmers the dynamic decisions about whether and how to compress data?

%% file: rs/acknowledgement.tex
\section{Acknowledgement}

We thank Isaac Gelado and Michael Garland (NVIDIA) for their foundational internship collaboration, thoughtful discussions, and helpful manuscript feedback.